\documentclass[prb,showpacs,superscriptaddress,twocolumn]{revtex4-2}
\usepackage{amssymb,amsmath,xcolor,graphicx,psfrag,bm}

\usepackage{hyperref, cancel}

\def\Xint#1{\mathchoice
   {\XXint\displaystyle\textstyle{#1}}%
   {\XXint\textstyle\scriptstyle{#1}}%
   {\XXint\scriptstyle\scriptscriptstyle{#1}}%
   {\XXint\scriptscriptstyle\scriptscriptstyle{#1}}%
   \!\int}
\def\XXint#1#2#3{{\setbox0=\hbox{$#1{#2#3}{\int}$}
     \vcenter{\hbox{$#2#3$}}\kern-.5\wd0}}

\def\dashint{\Xint-}

\begin{document}

\newcommand{\jav}[1]{{#1}}

\title{Non-hermitian off-diagonal magnetic response of Dirac fermions}

\author{Roberta Zs\'ofia Kiss}
\affiliation{Department of Theoretical Physics, Institute of Physics, Budapest University of Technology and Economics, M\H uegyetem rkp. 3., H-1111 Budapest, Hungary}

\author{Doru Sticlet}
\affiliation{National Institute for R\&D of Isotopic and Molecular Technologies, 67-103 Donat, 400293 Cluj-Napoca, Romania}
\author{C\u{a}t\u{a}lin Pa\c{s}cu Moca}
%\email{mocap@uoradea.ro}
\affiliation{MTA-BME Quantum Dynamics and Correlations Research Group, Institute of Physics, Budapest University of Technology and Economics, 
  M\H uegyetem rkp. 3., H-1111 Budapest, Hungary}
\affiliation{Department  of  Physics,  University  of  Oradea,  410087,  Oradea,  Romania}

\author{Bal\'azs D\'ora}
\email{dora.balazs@ttk.bme.hu}
\affiliation{Department of Theoretical Physics, Institute of Physics, Budapest University of Technology and Economics, M\H uegyetem rkp. 3., H-1111 Budapest, Hungary}
\affiliation{MTA-BME Lend\"ulet Topology and Correlation Research Group, Budapest University of Technology and Economics, M\H uegyetem rkp. 3., H-1111 Budapest, Hungary}
%\author{Doru Sticlet}
%\email{doru.sticlet@itim-cj.ro}
%\affiliation{National Institute for R\&D of Isotopic and Molecular Technologies, 67-103 Donat, 400293 Cluj-Napoca, Romania}
%\author{C\u{a}t\u{a}lin Pa\c{s}cu Moca}
%\email{mocap@uoradea.ro}
%\affiliation{MTA-BME Quantum Dynamics and Correlations Research Group, Institute of Physics, Budapest University of Technology and Economics, 
% Budafoki ut 8., H-1111 Budapest, Hungary}
%\affiliation{Department  of  Physics,  University  of  Oradea,  410087,  Oradea,  Romania}
%\noaffiliation

\date{\today}

\begin{abstract}
We perform a comparative study for the magnetization dynamics within linear response theory of one and two dimensional massive 
Dirac electrons, after switching on either a real (hermitian) or an imaginary (non-hermitian) magnetic field.
While hermitian dc magnetic fields polarize the spins in the direction of the external magnetic field, non-hermitian magnetic fields
induce only off diagonal response. 
An imaginary dc magnetic field perpendicular to the mass term induces finite magnetization in the third direction only according to the right hand rule.
This can be understood by analyzing the non-hermitian equation of motion of the spin, which 
 becomes analogous to a classical particle in crossed electric and magnetic fields. Therein, the spin expectation value, the
  mass term and imaginary magnetic field play the role of the classical momentum, magnetic and electric field, respectively. The latter two create a drift velocity 
perpendicular to them,  which gives rise to the off-diagonal component of the dc spin susceptibility, similarly to how the Hall effect develops in the classical description.

\end{abstract}

\maketitle

\section{Introduction}

With the advent of graphene and topological insulators\cite{CastroNeto2009,hasankane}, the Dirac equation has been essentially rediscovered in condensed matter physics, giving rise the
plethora of interesting effects in various dimensions and under several conditions. Due to the (pseudo)-spin structure in the Dirac equation\cite{song2015}, a variety of peculiar phenomena such as the   anomalous quantum Hall effect, electron chirality and Klein paradox
has been observed and this degree of freedom has been suggested to be useful for possible applications in spintronics\cite{he2019} and pseudospintronics\cite{pesin}.

In order for the (pseudo)-spin structure of the Dirac equation to be useful for applications, one needs to be able to control it with external field. While much is known about this within hermitian quantum mechanics, the effect of non-hermitian
external perturbations has been largely unexplored.  Non-hermitian systems have been extensively 
investigated\cite{gao2015,rotter,zeuner,Feng2014,hodaei,Bergholtz2021,ashidareview,ElGanainy2018,fruchart}
and present many unusual features, such as unidirectional invisibility\cite{Lin2011}, exceptional points\cite{heiss}, supersonic modes\cite{ashida2018}, 
the non-hermitian linear response 
theory\cite{linresp1,linresp2,linresp3} reveals unexpected features. These include measuring the anticommutators of observables instead of commutators, 
thus opening the door to access novel physical quantities experimentally as well as containing additional terms due to non-unitary dynamics of non-hermitian systems.

In order to shed light and investigate non-hermitian spin dynamics, we focus on the gapped hermitian Dirac equation 
in various dimensions and evaluate the real part of the magnetic 
susceptibility when the system is perturbed with a hermitian or a 
non-hermitian external magnetic field. We find that in response to the 
 hermitian magnetic fields, the spin susceptibility is a diagonal tensor
in the zero frequency limit, indicating that the induced magnetization always 
develops in the direction of the applied external field.
On the other hand, for imaginary magnetic field, the response is anisotropic and 
only the $xy$ component of the real part of the susceptibility is finite in the dc limit 
for a mass term in the $z$ direction. This can be understood by mapping the dynamics of the spins onto the 
classical Newton equation of a particle moving in a 
Lorentz force from magnetic field and an electric field. 
The former originates from the mass term while the latter stems from the non-hermitian magnetic field.
Within the context of the  classical Newton equation, these fields induce a drift velocity perpendicular to them and give rise to a 
finite momentum for the classical motion, which in turn is responsible
for the development of the Hall effect. In our case, in complete \emph{analogy} to the classical scenario, a finite
spin component is induced perpendicular to the mass term and imaginary magnetic field, 
according to the right hand rule.

\section{One dimensional Dirac equation}

We start with the one dimensional massive Dirac equation, whose Hamiltonian is
\begin{equation}
  H_0=vp\sigma_x+\Delta\sigma_z,
\end{equation}
where $\sigma$'s are Pauli matrices, denoting the (pseudo-)spin of the particles, $v$ is the Fermi velocity and $\Delta$ is the mass term, which couples to $\sigma_z$.
This is readily diagonalized to yield the spectrum $E_{\pm}=\pm\sqrt{(pv)^2+\Delta^2}$. 

We perform a comparative analysis and 
study the induced magnetization of the system in the long time limit.
The system is initially prepared in the ground state of $H_0$ at half filling with all $E_-$ energies occupied.
At  $t=0$ the system is perturbed with a weak real (hermitian) or imaginary (non-hermitian)
external magnetic field. 
%We apply a weak real and hermitian or imaginary and non-hermitian magnetic field at $t=0$ and 
To this end, we evaluate the \emph{real} part of the frequency dependent susceptibility
using the Kubo formula
from hermitian and non-hermitian linear response theory, whose
dc, $\omega\rightarrow 0$ limit is responsible for the value of the magnetization in the long time limit.

Let us also note that the particle current operator for the one dimensional Dirac equation is $v\sigma_x$, therefore
through measuring the magnetic response in the $x$ direction, implicitly the current correlation function is probed through the magnetoelectric effect\cite{hasankane}. 
However, for the other direction, 
$\sigma_y$ or $\sigma_z$ cannot be identified as a particle current operator. Nevertheless, all three Pauli matrices can be coupled to by magnetic fields.

\subsection{Hermitian magnetic field}

The magnetic response is evaluated using the Kubo formula for the spin susceptibility
\begin{equation}\label{sus}
\chi_{ij}(t,t')=-i\Theta(\tau)\langle [\sigma_i(\tau),\sigma_j]\rangle _0
%    \chi_{AB}(t,t')=-i\Theta(\tau)\langle [A(\tau),B]\rangle _0,
\end{equation}
in response to an external perturbation of the form $B\sigma_j$.
Here,  the expectation value is taken with respect to the ground state wavefunction of $H_0$, and $[A,B]$ denotes the commutator.
Here we introduced   $\tau=t-t'$  and $\sigma(\tau)=e^{i H_0 \tau}\sigma e^{-i H_0 \tau}$.
Then, the time dependence of the magnetization follows from
\begin{gather}
\langle\sigma_i(t)\rangle=\int_0^t\chi_{ij}(t,t')B(t')\mathrm{d}t'.
\end{gather}
Using the fact that the susceptibility is time translational invariant,  
$\chi_{ij}(t,t')=\chi_{ij}(t-t')$,  and performing the Fourier transformation, we get
\begin{gather}
\langle \sigma_i(\omega)\rangle=\chi_{ij}(\omega)B(\omega).
\label{chiomega}
\end{gather}
Since our focus is mostly on the possible finite magnetization in the long time limit, we need 
to evaluate Re$\chi(\omega\rightarrow 0)$, whose non-vanishing value
would signal finite magnetic response to a static magnetic field. 
Using Appendix \ref{appb}, the dc, $\omega\rightarrow 0$ limit of the real part of the magnetic susceptibilities are 
\begin{gather}\label{sus1}
    \underline{\underline{\chi}}(0)=\frac{1}{\pi v}
    \begin{pmatrix}
    |\mathrm{sgn}(\Delta)| & 0 & 0 \\
    0 & \ln\frac{2W}{|\Delta|} & 0 \\
    0 & 0 & \ln\frac{2W}{|\Delta|}
    \end{pmatrix},
\end{gather}
which is a diagonal matrix and  $W$ represents the high energy cutoff. This implies 
that a constant magnetization in the long time limit develops only in 
the direction of the applied external magnetic field.

\begin{figure}[t]
\centering
\includegraphics[width=7cm]{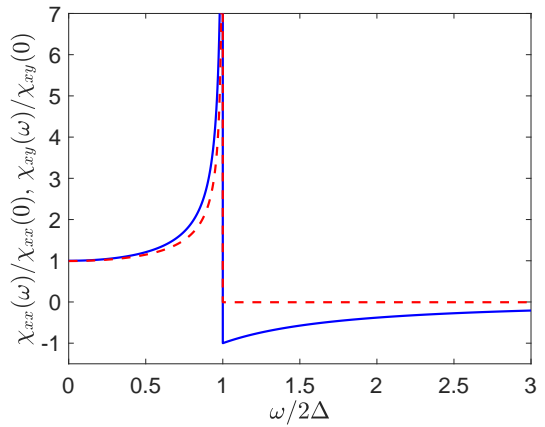}
\caption{The real part of the hermitian susceptibility (blue solid line), $\chi_{xx}$ and the non-hermitian one (red dashed line),
$\chi_{xy}$ is visualized for the one dimensional Dirac equation.}
\label{chi1d}
\end{figure}

\subsection{Non-hermitian magnetic field}

In the case of imaginary magnetic field, the full problem becomes 
non-hermitian and the weak perturbation is $iB\sigma_j$ which is apparently 
non-hermitian due to the $i$ prefactor.
The magnetic field is applied along the  $j=x$, $y$ or $z$, directions 
and $B$ is assumed real and denotes the strength of the non-hermitian magnetic field. 
Then, the operator to which the external perturbation couples to becomes
non-hermitian as $i\sigma_j$.
The corresponding real part of the susceptibility can be obtained from the non-hermitian Kubo formula\cite{linresp1,linresp2,linresp3} and 
Eq.~\eqref{chiomega} still applies for the non-hermitian setting.
In particular, in Ref. \onlinecite{linresp2}, some of the authors worked out the non-hermitian linear response theory and applied it to a non-hermitian
system (i.e.~tachyons) in the presence of hermitian perturbation (vector potential). In contrast to that, we now focus on hermitian systems, namely the Dirac equation,
in the presence of a non-hermitian perturbation, an imaginary magnetic field.

Adapting this to the present case of hermitian initial system in non-hermitian, imaginary perturbation, we obtain
\begin{equation}\label{sus2}
    \chi_{ij}(t,t')=-\Theta(\tau)\left(\left\langle\left\{\sigma_i(\tau),\sigma_j\right\}\right\rangle_0-2\langle \sigma_i\rangle_0 \langle\sigma_j\rangle_0\right),
\end{equation}
where $\{A,B\}$ denotes the anticommutator, which arises instead of the commutator due to the non-hermitian operator $i\sigma_j$, which the external field couples to.
After some straightforward algebra (see Appendix \ref{sec:Appendix})
and performing the Fourier transformation to frequency space, the momentum integrals of the non-zero elements for the real parts are evaluated as
\begin{gather}
%    \chi_{xx}(\omega)=-\frac{ \Delta^2}{2}\int_{-\infty}^{\infty}\mathrm{d}p\frac{1}{E^2}\Big(\delta(2E-\omega)+\delta(2E+\omega)\Big)=
\chi_{xx}(\omega)=-\frac{ \Delta^2}{2}\int\mathrm{d}p\frac{1}{E^2}\delta(2E-|\omega|)=\nonumber\\
=    -\frac{|\Delta|}{\sqrt{\frac{\omega^2}{4\Delta^2}-1}|\omega| v}\Theta(\omega^2-4\Delta^2),
\end{gather}

\begin{gather}
   \chi_{xy}(\omega)=\frac{2\Delta}{\pi}\dashint\mathrm{d}p\frac{1}{4E^2-\omega^2}=\nonumber\\
%   -\frac{\mathrm{sgn}(\Delta)}{\pi v }\times\nonumber\\
%\times\textmd{Re}\left(\frac{1}{\sqrt{\left(\frac{\omega}{2\Delta}\right)^2-1}}~ \mathrm{atanh}\left(\frac{W}{|\Delta| \sqrt{\left(\frac{\omega}{2\Delta}\right)^2-1}}\right)\right),
=\frac{\mathrm{sgn}(\Delta)}{2 v }\frac{\Theta(4\Delta^2-\omega^2)}{\sqrt{1-\left(\frac{\omega}{2\Delta}\right)^2}},
\end{gather}
\begin{gather}
%    \chi_{yy}(\omega)=-\frac{1}{2}\int_{-\infty}^{\infty}\mathrm{d}p\Big(\delta(2E-\omega)+\delta(2E+\omega)\Big)=\nonumber\\
\chi_{yy}(\omega)=-\frac{1}{2}\int\mathrm{d}p\delta(2E-|\omega|)=\nonumber\\
    =-\frac{1}{2\sqrt{1-\frac{4\Delta^2}{\omega^2}} v}\Theta(\omega^2-4\Delta^2),
\end{gather}
and
\begin{gather}
%    \chi_{zz}(\omega)=-\frac{1}{2}\int\mathrm{d}p\frac{(vp)^2}{E^2}\Big(\delta(2E-\omega)+\delta(2E+\omega)\Big)=\nonumber\\
\chi_{zz}(\omega)=-\frac{1}{2}\int_{-\infty}^{\infty}\mathrm{d}p\frac{(vp)^2}{E^2}\delta(2E-|\omega|)=\nonumber\\
=-\frac{1}{2v}\sqrt{1-\frac{4\Delta^2}{\omega^2}}\Theta(\omega^2-4\Delta^2).
\end{gather}
The only component which does not exhibit gapped behaviour is $\chi_{xy}$, whose frequency dependence is plotted in Fig. \ref{chi1d}.
The dc limit of the real part of these susceptibilities is evaluated as
\begin{gather}
    \underline{\underline{\chi}}(0)=\frac{\mathrm{sgn}(\Delta)}{2 v}
    \begin{pmatrix}
    0 & 1 & 0 \\
    -1 & 0 & 0 \\
    0 & 0 & 0 
    \end{pmatrix},
\label{od1}
\end{gather}
which is an off-diagonal matrix, and the only finite elements follows the "right hand rule", namely that the induced magnetization in the long time limit is perpendicular to 
both the mass term ($z$ direction in the present case) and the direction of the applied imaginary magnetic field. This is explained in Sec. \ref{lf}.

\section{Two dimensional Dirac equation}

The two dimensional gapped Dirac equation is written as
\begin{equation}
    H_0=v_xp_x\sigma_x+v_yp_y\sigma_y+\Delta\sigma_z,
\label{dirac2d}
\end{equation}
whose spectrum is $E_{\pm}=\pm\sqrt{(v_xp_x)^2+(v_yp_y)^2+\Delta^2}$, and the system is initially prepared in its ground state at half filling, i.e.~the $E_-$ energies are occupied.
Similarly to the one dimensional case, the particle current operators in the $x$ and $y$ directions are $v_x\sigma_x$ and $v_y\sigma_y$, while $\sigma_z$ cannot be
identified as a current.

\subsection{Hermitian magnetic field}

We use the conventional Kubo formula again from Eq.~\eqref{sus} and the Appendices. Eventually, 
in the $\omega \rightarrow 0$ limit, we get
\begin{gather}
    \underline{\underline{\chi}}(0)=\frac{W}{4\pi v_xv_y}
    \begin{pmatrix}
    1 & 0 & 0 \\
    0 & 1 & 0 \\
    0 & 0 & 2 
    \end{pmatrix},\label{herm2d}
\end{gather}
which is again a diagonal matrix, similarly to the one dimensional case.

\subsection{Non-hermitian magnetic field}

\begin{figure}[t]
\centering
\includegraphics[width=7cm]{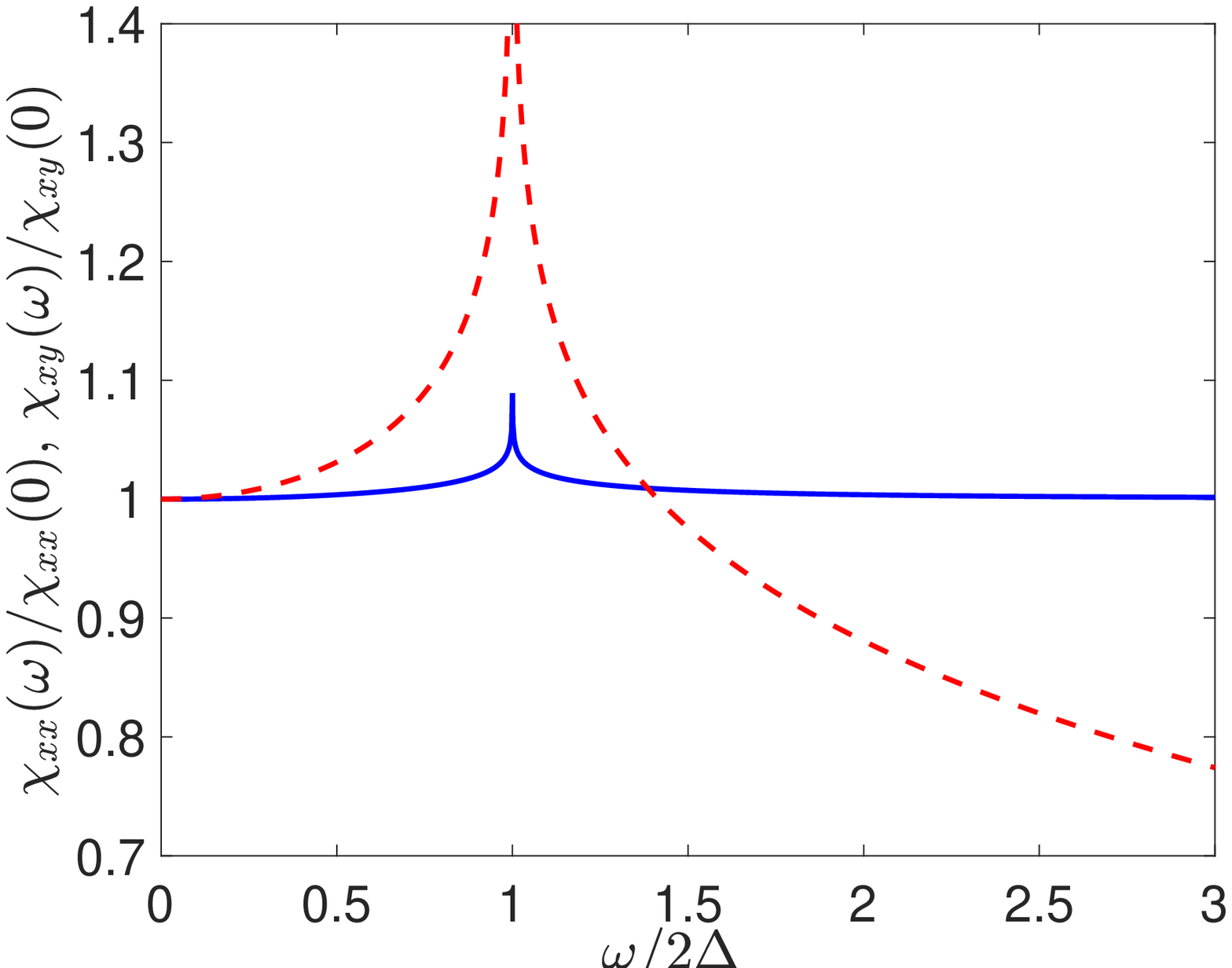}
\caption{The real part of the hermitian susceptibility (blue solid line), $\chi_{xx}$ and the non-hermitian one (red dashed line),
$\chi_{xy}$ is visualized for the two dimensional Dirac equation with $W/\Delta=100$.}
\label{chi2d}
\end{figure}

Similarly to the one dimensional case, we use the non-hermitian Kubo formula in Eq.~\eqref{sus2}.
The finite elements of the real part of the frequency dependent susceptibility from the Appendix are
\begin{gather}
    \chi_{xx}(\omega)=-\frac{1}{4\pi}\int \mathrm{d^2}p\frac{(v_yp_y)^2+\Delta^2}{E^2}
%\left(\delta(2E-\omega)+\delta(2E+\omega)\right)=
\delta(2E-|\omega|)=\nonumber\\
=
    -\frac{|\omega|}{16v_xv_y}\left(1+\frac{4\Delta^2}{\omega^2}\right)\Theta(\omega^2-4\Delta^2),
\label{2d1}
\end{gather}
\begin{gather}
   \chi_{xy}(\omega)=\frac{\Delta}{\pi^2}\dashint \mathrm{d^2}p\frac{1}{4E^2-\omega^2}=\nonumber\\
=
%   \frac{\Delta}{4 \pi v_xv_y}\textmd{Re}\left(\mathrm{ln}\left(\frac{\left(\frac{W}{\Delta}\right)^2+1-\left(\frac{\omega}{2\Delta}\right)^2}{1-\left(\frac{\omega}{2\Delta}\right)^2}\right)\right)
 \frac{\Delta}{4 \pi v_xv_y}\textmd{Re}\left(\mathrm{ln}\left(\frac{\left(\frac{W}{\Delta}\right)^2}{1-\left(\frac{\omega}{2\Delta}\right)^2}+1\right)\right),
\label{2d2}
\end{gather}
\jav{$\chi_{yy}(\omega)=\chi_{xx}(\omega)$} and
\begin{gather}
    \chi_{zz}(\omega)=-\frac{1}{4\pi}\int \mathrm{d^2}p\frac{(v_xp_x)^2+(v_yp_y)^2}{E^2}
%\left(\delta(2E-\omega)+\delta(2E+\omega)\right)=
\delta(2E-|\omega|)=\nonumber\\
=
    -\frac{|\omega|}{8v_xv_y}\left(1-\frac{4\Delta^2}{\omega^2}\right)\Theta(\omega^2-4\Delta^2).
\label{2d3}
\end{gather}
While the diagonal components are gapped, the only non-vanishing off-diagonal element, $\chi_{xy}$ is cutoff dependent, which is expected to dominate over the 
additional frequency dependence. This is shown in Fig. \ref{chi2d}.

After taking the $\omega \rightarrow 0$ limit, we find that the real part of the magnetic susceptibility is off-diagonal, similarly to the one dimensional case as
\begin{gather}
    \underline{\underline{\chi}}(0)=\frac{\Delta}{2\pi v_xv_y}\ln\left(\frac{W}{|\Delta|}\right)
    \begin{pmatrix}
    0 & 1 & 0 \\
    -1 & 0 & 0 \\
    0 & 0 & 0
    \end{pmatrix}.
\label{od2}
\end{gather}
The ensuing structure of Eq.~\eqref{od2} is explained in Sec. \ref{lf}.

\section{Three dimensional Dirac-Weyl equation}
\label{dw3d}

The three dimensional Dirac-Weyl equation is written as
\begin{equation}
    H_0=v_xp_x\sigma_x+v_yp_y\sigma_y+v_zp_z\sigma_z,
\label{dirac3d}
\end{equation}
whose spectrum is $E_{\pm}=\pm\sqrt{(v_xp_x)^2+(v_yp_y)^2+(v_zp_z)^2}$. This equation cannot be gapped out, a 
term of the form $\Delta\sigma_z$ does not open a gap in the spectrum.

\jav{In order to calculate the corresponding susceptibilities, we realize that the three dimensional Dirac-Weyl 
equation in Eq. \eqref{dirac3d} can be obtained from the two dimensional gapped Dirac equation 
from Eq. \eqref{dirac2d} by replacing $\Delta$ with $v_zp_z$. 
Consequently, the frequency dependent susceptibility of the former is obtained from that of the latter in Eqs. \eqref{app2dherm} and \eqref{app2dnonherm}
after the same 
$\Delta\rightarrow v_zp_z$ replacement and integrating over $p_z$. 
When moving to the dc limit, the corresponding dc susceptibility is obtained by performing the same
replacement in Eqs. \eqref{herm2d} and \eqref{od2}. 
For the hermitian case, the dc susceptibility is finite and independent from $\Delta$ in Eq. \eqref{herm2d} for the two dimensional case.
By moving to the Dirac-Weyl case with the $\Delta\rightarrow v_zp_z$ replacement and integrating over $p_z$, the dc limit of the hermitian
susceptibility is finite as $\chi(0)\sim W^2$, as already identified for three dimensional Dirac semimetals in Ref. \onlinecite{ominato}.}

\jav{In contrast, Eq. \eqref{od2} is odd in $\Delta$ for the non-hermitian case in two dimensions. Due to this, by moving into the Dirac-Weyl case with
$\Delta\rightarrow v_zp_z$ change and momentum integration, it
 will vanish.
Therefore, we find that  the dc limit of the non-hermitian susceptibilities for Dirac-Weyl systems are zero
as 
\begin{gather}
\chi(0)=0. 
\label{nh3d}
\end{gather}
}
No finite magnetization can be induced by a
static, imaginary magnetic field at half filling (when the $E_-$ band is filled.

\jav{By applying the same procedure to Eqs. \eqref{2d1}, \eqref{2d2} and \eqref{2d3}, we obtain the frequency dependent non-hermitian susceptibilities as well.
The off-diagonal components vanish (Eq. \eqref{2d2} is odd in $\Delta$) and the diagonals are equals to each other as $\chi(\omega)=-\omega^2/24\pi v_xv_yv_z$.}

\section{Equation of motion for the spin\label{lf}}

The off-diagonal nature of the real part of the non-hermitian dc magnetic susceptibility can be understood by inspecting the equation of motion for the spins\cite{botet2018}.
We consider the full non-hermitian Hamiltonian with $H=\left({\bf A_p}+i\bf{B}\right)\cdot \bm{\sigma}$, where
 ${\bf A_p}=(vp,0,\Delta)$ for the one dimensional case
and
 ${\bf A_p}=(v_xp_x,v_yp_y,\Delta)$ for the two dimensional case and $i\bf B$ 
denotes the imaginary time independent magnetic field, 
that is switched on at $t=0$. The expectation value of the spin for a given momentum 
is evaluated from\cite{daley,ashidareview,carmichael,graefe2008}
\begin{gather}
\langle {\bm\sigma_{\bf p}}(t)\rangle=\frac{\langle \Psi_{\bf p}|e^{iH^+t}{\bm\sigma}e^{-iHt}|\Psi_{\bf p}\rangle}
{\langle \Psi_{\bf p}|e^{iH^+t}e^{-iHt}|\Psi_{\bf p}\rangle},
\end{gather}
and the system starts from lowest energy eigenstate, $\Psi_{\bf p}$, of ${\bf A_p}\cdot \bm{\sigma}$.
The equation of motion for the spin for a given momentum $\bf p$ reads as
\begin{gather}
\partial_t \langle {\bm\sigma_{\bf p}}(t)\rangle=2{\bf A_p}\times  \langle {\bm\sigma_{\bf p}}(t)\rangle-2{\bf B}+2\langle {\bm\sigma_{\bf p}}(t)\rangle \left[\langle {\bm\sigma_{\bf p}}(t)\rangle\cdot {\bf B}\right],
\label{lorentz}
\end{gather}
which resembles closely to the Newton's equation of a classical particle in a crossed electric and magnetic field. Here, $\langle {\bm\sigma_{\bf p}}(t)\rangle$ represents the classical momentum, $\bf A_p$ plays the role of the magnetic field and the first term on the r.h.s. of Eq.~\eqref{lorentz} represents the Lorentz force,
$\bf B$ represents an electric field and the last term is the relativistic correction. 

In a crossed electric and magnetic field, the particle experiences a drift velocity for the guiding center\cite{NORTHROP}, 
which is perpendicular to both the electric and magnetic field. In this case, this effective "drift velocity" points towards $\bf B\times A_p$, 
and after averaging over momentum, in order to get the total spin as $\sum_{\bf p} \langle {\bm\sigma_{\bf p}}(t)\rangle$, it becomes 
perpendicular to both the $z$ direction (the direction of the mass term, which does not average out) and the direction of the imaginary magnetic field. 
This results in an effective magnetization in the perpendicular direction, in agreement with Eqs. \eqref{od1} and \eqref{od2}, similarly
 to how the Hall-effect develops in the classical case.

\jav{The equation of motion method also allows us to analyze the three dimensional Dirac-Weyl case with ${\bf A_p}=(v_xp_x,v_yp_y,v_zp_z)$, which
can be obtained from the two dimensional case after the $\Delta \rightarrow v_zp_z$ change and integration over $p_z$, similarly to Sec. \ref{dw3d}.
 For a given 
fix $p_z$, there will be a 
finite magnetization
developing perpendicular to both the non-hermitian magnetic field and $z$ direction, analogously to the two dimensional case. However, this will be compensated 
exactly by the contribution of the 
$-p_z$ term, which arises from the momentum integration. This will cancel the magnetization from the $+p_z$ term and yield Eq. \eqref{nh3d}.}

\section{Experimental possibilities}

\jav{In terms of experimental realization, various forms of the Dirac equation in various spatial dimensions 
have already been realized\cite{CastroNeto2009,Gerritsma2010,tarruell,Lamata2007,Lee2015,Zhen2015,Song2020} both in condensed matter
and cold atomic systems as well as in photonic crystals.
In an open quantum system, interacting with its environment, 
the non-hermitian term (i.e.~the imaginary magnetic field) arises 
from an effective Lindblad equation without the recycling term\cite{daley,carmichael}
through continuous monitoring of the system and postselection\cite{daley,carmichael},
using jump operators for bonds\cite{ashida2018,gongprx,takasu}. 
This results in the appropriate imaginary magnetic fields for the one, 
two and three dimensional Dirac equations.
By preparing the system in a given initial state with the $E_-$ band filled and coupling it weakly to environment, the (pseudo-)magnetization 
in a given direction can be measured, which would directly yield the calculated susceptibilities.

One can also profit from the recent non-hermitian realization of spin-orbit coupled fermions\cite{ren2022}. By considering weak spin dependent
$^{173}$Yb atom losses, a non-hermitian magnetic field can be engineered. By monitoring the ensuing time dependent spin profile, 
the non-hermitian spin susceptibility is directly
accessible after Fourier transformation to frequency space.
 
Photonic waveguides\cite{zeuner,Song2020,Zhen2015,longhi2010} are also used to emulate the non-hermitian Dirac equation, with the complex refractive index due 
to losses 
representing the non-hermitian
term. Weak imaginary magnetic fields are created by weak losses, and by measuring light propagation across the experimental setup
yields the non-hermitian pseudo-magnetic susceptibilities within the validity range of our linear response calculations.

Additionally, single photon interferometry is also available to realize Dirac equation in the presence of non-hermitian
magnetic fields\cite{PRXdora} directly in momentum space. By preparing the system in the lower energy band and controlling the time evolution
in the presence of weak non-hermitian terms by a variety of optical elements, the time dependent magnetization can be obtained.}

\section{Conclusions}

We studied the magnetization dynamics in terms of the real part of the frequency dependent spin susceptibility of one, two and three dimensional gapped Dirac electrons, in response to hermitian or non-hermitian magnetic fields. By focusing on the long time limit of the magnetization, we find that
a hermitian magnetic field induces diagonal response and the ensuing spin expectation value points in the direction of the external perturbation.
In sharp contrast, a non-hermitian magnetic field triggers off-diagonal response according to the right hand rule: a constant magnetization develops in the direction perpendicular to both the direction of the mass term and that of the non-hermitian magnetic field.
This is understood by mapping the equation of motion of the spin to a Newton equation of a classical particle in electric and magnetic fields, the latter giving rise to the Lorentz force. In the classical case, a finite drift velocity develops perpendicular to both the electric and magnetic fields.
Analogously for the spin, a finite spin expectation value is only expected perpendicular to both the mass term and the imaginary magnetic field. Our results could be useful for further manipulation of the spin of Dirac particles in open quantum systems and in dissipative environment.

\begin{acknowledgments}
Useful discussions with Ferenc Simon, J\'anos Asb\'oth are gratefully acknowledged.
This research is supported by the National Research, Development and Innovation Office - NKFIH  within the Quantum Technology National 
Excellence Program (Project No.~2017-1.2.1-NKP-2017-00001), K142179, K134437, by the BME-Nanotechnology FIKP grant (BME FIKP-NAT), and 
by a grant of the Ministry of Research, Innovation and Digitization, CNCS/CCCDI-UEFISCDI, under Projects No.~PN-III-P4-ID-PCE-2020-0277 and PN-III-P1-1.1-TE-2019-0423.
\end{acknowledgments}

\bibliographystyle{apsrev}
\bibliography{wboson1}

\appendix

\section{Calculation of the susceptibility}\label{sec:Appendix}
\begin{widetext}
In one dimension with hermitian magnetic field, we get
\begin{gather}\label{er1}
    \underline{\underline{\chi}}(t,t')=
    \frac{1}{2\pi}\int \mathrm{d}p\frac{2}{E^2}\begin{pmatrix}
       \Delta^2\sin{(2E\tau)} & \Delta E \cos{(2E\tau)} & -\Delta vp \sin{(2E\tau)} \\
        -\Delta E\cos{(2E\tau)} & E^2\sin{(2E\tau)} &  vp E\cos{(2E\tau)} \\
        -\Delta vp\sin{(2E\tau)} & - vp E\cos{(2E\tau)} &  (vp)^2 \sin{(2E\tau)}
    \end{pmatrix}\Theta(\tau).
\end{gather}
where $E=\sqrt{(vp)^2+\Delta^2}$.

In one dimension with imaginary magnetic field, we obtain
\begin{gather}\label{er2}
    \underline{\underline{\chi}}(t,t')=
    \frac{1}{2\pi}\int \mathrm{d}p\frac{2}{E^2}
    \begin{pmatrix}
       -\Delta^2\cos{(2E\tau)} & \Delta E \sin{(2E\tau)} & \Delta vp \cos{(2E\tau)} \\
        -\Delta E\sin{(2E\tau)} & -E^2\cos{(2E\tau)} &  vp E\sin{(2E\tau)} \\
        \Delta vp\cos{(2E\tau)} & -vp E\sin{(2E\tau)} & -(vp)^2 \cos{(2E\tau)}
    \end{pmatrix}\Theta(\tau).
\end{gather}

In two dimensions with hermitian magnetic field, the susceptibility is
\begin{gather}\label{er3}
    \underline{\underline{\chi}}(t,t')=
    \frac{1}{4\pi^2}\int \mathrm{d}p_x\mathrm{d}p_y\frac{2}{E^2}
    \begin{pmatrix}
       ((v_yp_y)^2+\Delta^2) & -(v_yp_y)(v_xp_x) & -(v_xp_x)\Delta \\
        -(v_yp_y)(v_xp_x) & ((v_xp_x)^2+\Delta^2) &  -(v_yp_y)\Delta \\
        -(v_xp_x)\Delta & -(v_yp_y)\Delta & ((v_xp_x)^2+(v_yp_y)^2)
    \end{pmatrix} \sin(2E\tau)\Theta(\tau)\nonumber \\
    +\frac{1}{4\pi^2}\int \mathrm{d}p_x\mathrm{d}p_y\frac{2}{E}
    \begin{pmatrix}
       0 & \Delta & -(v_yp_y) \\
        -\Delta & 0 &  (v_xp_x) \\
        (v_yp_y) & -(v_xp_x) & 0
    \end{pmatrix}\cos(2E\tau)\Theta(\tau),
\label{app2dherm}
\end{gather}
where $E=\pm\sqrt{(v_xp_x)^2+(v_yp_y)^2+\Delta^2}$.
The two dimensional case with imaginary magnetic field yields
\begin{gather}\label{er4}
    \underline{\underline{\chi}}(t,t')=
    \frac{1}{4\pi^2}\int \mathrm{d}p_x\mathrm{d}p_y\frac{2}{E^2}
    \begin{pmatrix}
       -((v_yp_y)^2+\Delta^2) & (v_yp_y)(v_xp_x) & (v_xp_x)\Delta \\
        (v_yp_y)(v_xp_x) & -((v_xp_x)^2+\Delta^2) &  (v_yp_y)\Delta \\
        (v_xp_x)\Delta & (v_yp_y)\Delta & -((v_xp_x)^2+(v_yp_y)^2)
    \end{pmatrix} \cos(2E\tau)\Theta(\tau)\nonumber \\
    +\frac{1}{4\pi^2}\int \mathrm{d}p_x\mathrm{d}p_y\frac{2}{E}
    \begin{pmatrix}
       0 & \Delta & -(v_yp_y) \\
        -\Delta & 0 &  (v_xp_x) \\
        (v_yp_y) & -(v_xp_x) & 0
    \end{pmatrix}\sin(2E\tau)\Theta(\tau).
\label{app2dnonherm}
\end{gather}
\end{widetext}
The susceptibility of three dimensional Dirac-Weyl fermions follows from Eqs. \eqref{app2dherm} and \eqref{app2dnonherm} after replacing $\Delta$ with $v_zp_z$.

\section{Frequency dependent susceptibilities for hermitian magnetic field}\label{appb}

\subsection{One dimension}

\jav{Based on  Appendix \ref{sec:Appendix}, the non-vanishing components of the real part of the susceptibility (with $W$ the high energy cutoff) are
\begin{gather}
    \chi_{xx}(\omega)=\frac{2 \Delta^2}{\pi}\dashint\mathrm{d}p\frac{1}{E\left(4E^2-\omega^2\right)}=
    -\frac{1}{\pi v |\frac{\omega}{2\Delta}|}\times\nonumber\\
\times\textmd{Re}\left(\frac{1}{\sqrt{\left(\frac{\omega}{2\Delta}\right)^2-1}}~ \mathrm{atanh}
\left(\frac{1}{\sqrt{1-\left(\frac{2\Delta}{\omega}\right)^2}}\right)\right)
\end{gather}
with $E=\sqrt{(vp)^2+\Delta^2}$ and $\dashint$ denoting Cauchy’s principal value of an integral,
\begin{gather}
%    \chi_{xy}(\omega)=\frac{\Delta}{2}\int_{-\infty}^{\infty}\mathrm{d}p\frac{1}{E}\Big(\delta(2E-\omega)+\delta(2E+\omega) \Big)=
\chi_{xy}(\omega)=\frac{\Delta}{2}\int\mathrm{d}p\frac{1}{E}\delta(2E-|\omega|)=\nonumber\\
=
    \frac{\Delta}{\sqrt{\omega^2-4\Delta^2}v}\Theta(\omega^2-4\Delta^2),
\end{gather}

\begin{gather}
    \chi_{yy}(\omega)=\frac{2}{\pi}\dashint\mathrm{d}p\frac{E}{4E^2-\omega^2}=
    \frac{1}{\pi v}\left(\ln\left(\frac{2W}{|\Delta|}\right)-\right.\nonumber\\
-\left.\textmd{Re}\left(\frac{1}{\sqrt{1-\left(\frac{2\Delta}{\omega}\right)^2}}
\mathrm{atanh}\left(\frac{1}{\sqrt{1-\left(\frac{2\Delta}{\omega}\right)^2}}\right)\right)\right),
\end{gather}
and
\begin{gather}
    \chi_{zz}(\omega)=\frac{2}{\pi}\dashint\mathrm{d}p\frac{(vp)^2}{E(4E^2-\omega^2)}=
    \frac{1}{\pi v}\left(\mathrm{ln}\left(\frac{2W}{|\Delta|}\right)-\right.\nonumber\\
-\left.\textmd{Re}\left(\sqrt{1-\left(\frac{2\Delta}{\omega}\right)^2}\mathrm{atanh}
\left(\frac{1}{\sqrt{1-\left(\frac{2\Delta}{\omega}\right)^2}}\right)\right)\right)
\end{gather}
In the continuum limit, the allowed momenta region is extended to infinity, 
but in order to make contact with the original model, a cutoff needs to be introduced for certain non-universal physical quantities.
The cutoff presence is rather natural in effective low energy theories\cite{giamarchi}, and accounts for
the finite bandwidth, which is present in the original tight binding Hamiltonian, and stems from the finite Brillouin zone.}

\jav{Among these components, $\chi_{xy}$ exhibits a gapped behaviour (i.e.~$|\omega|>2\Delta$ is required), while $\chi_{yy}$ and $\chi_{zz}$ are
dominated by the cutoff dependent term, which dominates over the additional
frequency dependences. The frequency dependence of the cutoff independent $\chi_{xx}$ is shown in Fig. \ref{chi1d}.}

\subsection{Two dimensions}

\jav{Based on  Appendix \ref{sec:Appendix}, we get
\begin{gather}
    \chi_{xx}(\omega)=\frac{1}{\pi^2}\dashint \mathrm{d^2}p\frac{(v_yp_y)^2+\Delta^2}{E\left(4E^2-\omega^2\right)}=
    \frac{(W-|\Delta|)}{4\pi v_xv_y}+\nonumber\\
+\frac{ |\omega|}{8\pi v_xv_y}\left(1+\left(\frac{2\Delta}{\omega}\right)^2\right)
\textmd{Re}\left(\mathrm{atanh}\left(\left|\frac{2\Delta}{\omega}\right|\right)\right),
\end{gather}
where $E=\pm\sqrt{(v_xp_x)^2+(v_yp_y)^2+\Delta^2}$, $\chi_{yy}(\omega)=\chi_{xx}(\omega)$ and
\begin{gather}
%    \chi_{xy}(\omega)=\frac{\Delta}{4\pi}\int_{-\infty}^{\infty} \mathrm{d}p_x \int_{-\infty}^{\infty} \mathrm{d}p_y\frac{1}{E}
%\Big(\delta(2E-\omega)+\delta(2E+\omega)\Big)=
 \chi_{xy}(\omega)=\frac{\Delta}{4\pi}\int \mathrm{d^2}p\frac{1}{E}\delta(2E-|\omega|)= \nonumber\\
=
  \frac{\Delta}{4v_xv_y}\Theta(\omega^2-4\Delta^2),
\end{gather}
and
\begin{gather}
    \chi_{zz}(\omega)=\frac{1}{\pi^2}\dashint\mathrm{d^2}p\frac{(v_xp_x)^2+(v_yp_y)^2}{E\left(4E^2-\omega^2\right)}=
    \frac{(W-|\Delta|)}{2\pi v_xv_y}+\nonumber\\
+\frac{ |\omega|}{4\pi v_xv_y}\left(1-\left(\frac{2\Delta}{\omega}\right)^2\right)
\textmd{Re}\left(\mathrm{atanh}\left(\left|\frac{2\Delta}{\omega}\right|\right)\right).
\end{gather}
The diagonal components scale with the cutoff, which is expected to overwhelm  the additional frequency dependences, while the off-diagonal 
piece, $\chi_{xy}$ exhibits
gapped behaviour and contributes only for $|\omega|>2\Delta$. We note that similar cutoff dependent
spin susceptibility, $\chi(\omega\rightarrow 0)\sim W$,  was observed experimentally for two dimensional  topological Dirac fermions\cite{zhao}.
Therein, by measuring the singular spin response of Dirac fermions, the effective cutoff was extracted from the experimental data in Fig. 3 
of \onlinecite{zhao}.
In addition, $\chi_{xx}(\omega)$ agrees with the dynamical current-current susceptibility of gapped graphene in Ref. \onlinecite{scholz}.}

\subsection{Three dimensions}

\jav{
By replacing $\Delta$ with $v_zp_z$ in Eq. \eqref{app2dherm} and performing the momentum integrals, we find that the resulting susceptibility is diagonal and the diagonal elements are equal as
\begin{gather}
\chi(\omega)=\frac{1}{3\pi^2 v_xv_yv_z}\left(W^2+\frac{\omega^2}{4}\ln\left(\frac{4W^2+\omega^2}{\omega^2}\right)\right).
\end{gather}
It is dominated by the first, frequency independent term $\sim W^2$.
}

\end{document}